\documentclass[12pt]{article}

\def\ifanon{\ifx43}   

\usepackage{amsmath}
\usepackage{amssymb}
\usepackage{jsrnft}
\usepackage{xcolor}
\usepackage{xspace}
\usepackage{url}

\newtheorem{proposition}{Proposition}

\def\U{{\mathbf U}}
\def\Pr{{\mathbf P}}

\def\T{{\cal T}}

\begin{document}

\baselineskip 24pt

\begin{center}
{\titlefont Football Group Draw Probabilities and Corrections}
\\ \medskip

\ifanon
\vskip 4cm
\else

by
\\ \smallskip
Gareth O. Roberts\footnote
{\baselineskip=12pt
\parskip=0pt
Department of Statistics,
University of Warwick, CV4 7AL, Coventry, U.K.
Email: {\tt Gareth.O.Roberts@warwick.ac.uk}.
}
\ and \
Jeffrey S. Rosenthal\footnote
{\baselineskip=12pt
Department of Statistics, University of Toronto,
Toronto, Ontario, Canada
\
M5S 3G3.
Email:
{\tt jeff@math.toronto.edu}.  Web: {\tt http://probability.ca/jeff/}
}
\\ \smallskip

\fi  

(Version of: \today)
\end{center}
\bigskip

\begin{quote}
\baselineskip=12pt
{\bf Abstract:} \
This paper considers the challenge of designing football group draw
mechanisms which have the uniform distribution over all valid draw
assignments, but are also entertaining, practical, and transparent. Although this
problem is trivial in completely symmetric problems, it becomes challenging
when there are draw constraints which are not exchangeable across each of the
competing teams, so that symmetry breaks down. We
explain how to simulate the FIFA Sequential Draw method, and compute the
non-uniformity of its draws by comparison to a uniform Rejection Sampler.
We then propose two practical methods of achieving the uniform
distribution while still using balls and bowls in a way which is suitable
for a televised draw.  The solutions can also be carried out interactively.
The general methodology we provide can readily be transported to different
competition draws and is not restricted to football events.
\end{quote}


\sect{Introduction}

Major football (soccer) tournaments such as the FIFA World
Cup, European Championships, and UEFA Champions League
hold public draws to
decide who plays whom. It is customary to employ one or more celebrities
to draw balls from pots to sequentially construct the draw, to
add excitement and increase
interest in the competition. However, such mechanisms can affect the
draw probabilities in unexpected ways.

This article will focus on the group draw for the
FIFA World Cup, although similar
ideas could be applied to other competitions. This
event involves teams from 32 different nations.
2022 was of particular interest to Canada, who qualified
for the first time since 1986.
Its group
draw had various restrictions, based upon geographical constraints,
about which assignments are permissible, leading to a complicated space
of potential draws.
In addition, the draw needed to take place
sequentially, to allow for public interest and transparency.
The statistical challenge, then, would be to simulate from the
uniform distribution on a non-symmetric high-dimensional space in a way
which is also sequential and entertaining.
We shall present several potential solutions that we have developed
to address this challenge;
they are available for interactive use~\cite{drawweb},
and have been reported on in the media~\cite{starart,cbcart}.
See also their simple informal descriptions in
Section~\ref{sec-informal} below.

The methodology we develop is quite general and is readily transportable to different
football competition draws, and indeed to other sports which carry out public
draws of this kind (for example the 2021 World Men's Handball Championships, \cite{2021Hand}). We will focus on the FIFA World Cup
given its global appeal.

Sequential simulation of complex probability distributions has a long
history of use in statistics (e.g.\ particle filtering, nested sampling, rare
event simulation, simulated annealing, etc.). This paper considers a
rather different application of these techniques, which nevertheless require 
statistical thinking, design, and analyses. We will thus utilise key state of the art
techniques from the Computational Statistics literature such as stratified resampling
for sequential Monte Carlo \cite{kitagawa1996monte}, and retrospective simulation  for the simulation of
distributions with intractable probabilities, see e. g. \cite{retrospect}.

\subsect{The 2022 FIFA World Cup group draw}
\labelss{sec-wctt}

\def\Eu{{\bf Eu}}
\def\SA{{\bf SA}}
\def\NA{{\bf NA}}
\def\As{{\bf As}}
\def\Af{{\bf Af}}
\def\Oc{{\bf Oc}}

The 2022 FIFA World Cup took place in November/December
2022, in Qatar.  It involved 32 national teams --
31 who qualified through competition, plus
the host team Qatar who qualified automatically.
These 32 teams needed to be partitioned
into 8 groups of 4 teams each, who would all play each other in
the group stage of the competition.

The group assignments were subject to various restrictions as follows.
The 32 teams were first partitioned into
4 seeded pots, with Pot~1 consisting of the hosts Qatar plus
the seven most highly ranked teams (according
to the official FIFA national team rankings), Pot~2 consisting of
the next 8 highest ranked teams, and so on (with the proviso that at the time there were three
places still to be decided via various playoffs, and each of these
undetermined slots were placed in Pot~4).
In addition,
each team is affiliated to one of the 6 continental federation regions:
UEFA (Europe; 13 teams; henceforth \Eu),
CONMEBOL (South America; 4 or 5 teams; henceforth \SA),
CONCACAF (North and Central America; 3 or 4 teams; henceforth \NA),
AFC (Asia; 5 or 6 teams; henceforth \As),
CAF (Africa; 5 teams; henceforth \Af),
and OFC (Oceania; 0 or 1 team; henceforth \Oc).
The assigned pots were as follows:

{\bf Pot~1:}  Qatar[\As], Belgium[\Eu], Brazil[\SA], France[\Eu],
                Argentina[\SA], England[\Eu], Portugal[\Eu], Spain[\Eu].

{\bf Pot~2:}  Denmark[\Eu], Netherlands[\Eu], Germany[\Eu], Switzerland[\Eu],
                Croatia[\Eu], Mexico[\NA], USA[\NA], Uruguay[\SA].

{\bf Pot~3:}  Iran[\As], Serbia[\Eu], Japan[\As], Senegal[\Af],
                Tunisia[\Af], Poland[\Eu], KoreaRep[\As], Morocco[\Af].

{\bf Pot~4:}  Wales/Scot/Ukr[\Eu], Peru/UAE/Au[\SA,\As],
CostaRica/NZ[\NA,\Oc], SaudiArabia[\As],
                Cameroon[\Af], Ecuador[\SA], Canada[\NA], Ghana[\Af].

\medskip\noindent
(The reason for the uncertainty in three of the team names
in Pot~4 is that, due to
delays caused by Covid-19 and the war in Ukraine,
not all teams had been finalised by the time of the draw,
so placeholders were used.  Two of the placeholders
corresponded to two different potential regions, so they had
to satisfy the geographical constraints
for {\it both} of the corresponding regions.)

In terms of these specifications, the restrictions on group
formations were that
each group needed to include one team
from each of the 4 pots, and furthermore include
either 1 or 2 teams from \Eu\
plus either 0 or 1 teams from each of the other regions.

The FIFA group draw, held on 1 April 2022, then proceeded as follows~\cite{fifa2022}.
First, the host team Qatar was automatically placed in Group~A.
Then, the remaining teams from Pot~1
were selected one at a time, uniformly at random, and each
placed into the next group from B through H.
Then, the teams from Pot~2
were selected one at a time, uniformly at random,
and assigned to the next available group,
i.e.\ the first group which would not cause a conflict
with any of the geographical restrictions (either immediately in
the group where they were placed,
or subsequently by making it impossible to validly
fill in the remainder of the draw).
This procedure was then repeated with Pot~3,
and then with Pot~4 to conclude the draw.
Each random selection was performed by a celebrity footballer,
who picked a ball uniformly at random from a round bowl,
and opened it to reveal the chosen team.
(Each team was also randomly
assigned a ``position'' within their group, to determine
the order in which the matches would be played,
but we do not consider that issue here.)

\def\NumbDraws{$5.9\times 10^{14}$}
\def\ratio{560}

Without the geographical constraints,
just drawing the teams in random order from the four pots in sequence,
there would be $7! \times (8!)^3 \doteq 3.3 \times 10^{17}$
possible draws that could be produced. We shall
see below that about 1 in \ratio\ of these unconstrained draws satisfy the
geographical constraints.  Hence,
the number of valid draws is approximately \NumbDraws.
A {\em uniform} draw is one for which all 
valid draws
have an equal chance of materialising.
This article will explore the non-uniformity
in the FIFA 2022 draw procedure, and also propose various
methods of refining the draw to achieve complete uniformity.

\subsect{Previous literature}

Unfairness of sporting rules is currently an active area of research in the Operations Research literature, see e.g. \ 
\cite{kendall2017sports,lenten2021scholarly,csato2021tournament}, and
hidden biases in draw mechanisms provide good examples of this. 
It has long been known that sequential draws such as those adopted by FIFA
and UEFA led to non-uniform probabilities, see e.g.\
\cite{jones1990world,rathgeber2007germany,klossner2013odd,guyon2015rethinking,csato2021fairness}. Various papers
have looked at different mechanisms for carrying out different sorts of
sequential draws; see e.g.\ \cite{csato2021fairness}. 
Much of the
literature has focused on how to obtain a {\em balanced} draw rather than
a uniform one, i.e.\ trying to make each group roughly equal in strength
(see e.g.~\cite{Guyon18,csato2022unknown,cea2020analytics,laliena2019fair}).
This literature has been successful
in influencing rules for football draws.
Indeed, balance is what inspired FIFA
to create the seeded pots based on world rankings rather than continental
affiliation, for both the 2018 and 2022 World Cup group draws.
However seeking balanceness by adjusting the constraints within the draw
 is a separate question from the question of obtaining uniformity without
 changing constraints (which is the focus of this paper).

Other papers consider other related issues, such as
how teams' incentives to perform well are affected by
tournament designs \cite{guyon2021choose}
and by group draws \cite{csato2022incentive}.

One might wonder whether non-uniformity really matters in practice, or whether it is merely of theoretical interest.
However a study based on the UEFA Champions League \cite[p.~262]{klossner2013odd} shows that even
small probability differences can translate into quite substantial financial differences in expected revenues for different clubs. For the 2018 FIFA World Cup, \cite{csato2021fairness} finds that small biases can also significantly affect nations' progression probabilities beyond the initial group stage.

Various papers have proposed solutions to the non-uniformity problem.
The paper \cite{boczon2018goals}
produces a sequential procedure which is {\it closer} to uniform than
existing methods, but is still not completely uniform.
The work \cite{guyon2015rethinking} proposes three
innovative solutions, each of which are truly transparent using
only balls and bowls (with no multiple bowls) and very promising, but have
certain acknowledged limitations:
their Suggestion~1 does not lead to a uniform draw;
their Suggestion~2 is uniform but
requires intractable preliminary computation to list all 
possible continental distributions;
and their Suggestion~3 adds an additional ``S-curve-type'' constraint
which increases group balance but significantly decreases the number of
admissible draws, so it produced a tractable transparent
draw which is uniform but only on a different (much smaller) draw set.
Also, the paper \cite{klossner2013odd} briefly
postulates a possible Markov chain procedure (specifically an exclusive
Gibbs sampler) for the UEFA Champions League draw.  However, none of these
methods respect the desired sequential nature of the draw in a practically feasible way
while also achieving perfect uniformity over all draws satisfying the
specified FIFA constraints, which is our goal here.

\subsect{Summary of main contributions and structure of paper}
\labels{sec-maincontstr}

Our main contributions in this paper are
\begin{enumerate}
\item
We explicitly calculate biases inherent in the 2022 FIFA World Cup draw procedure.
\item
We propose three practically implementable solutions to completely remove bias. All of these methods could potentially be used by FIFA (and other sporting bodies) for improved future draws, no matter what constraints are imposed .
\item
We give prototype software to illustrate our solutions.
\item
We provide a user's guide to sporting bodies to suggest how these solutions could be implemented in ways which are fairly similar to current procedures.
\end{enumerate}

In Section \ref{sec-compare} we investigate the extent of the bias in the 2022 FIFA draw, comparing  a simulation of the FIFA draw mechanism with uniform draws which are achieved by rejection sampling. In Section \ref{sec-motiv}, we attempt to demystify this bias created by the FIFA mechanism via a simplified example and investigate how multiple ball procedures could potentially be utilised to circumvent the problem. The next three sections are devoted to our solutions: Section \ref{sec-swap} describes the Metropolis (swap) solution; Section \ref{sec-multballs} introduces our multiple ball solution, while a multiple rejection solution is provided in Section \ref{sec-multirej}.
Section~\ref{sec-informal} provides informal, non-technical descriptions
of our two most practical methods.
We conclude with a discussion of various related issues
in Section \ref{sec-disc}.

\sect{Comparing uniform and FIFA probabilities}
\labels{sec-compare}

Before proposing alternative solutions, we investigate the
extent to which the FIFA Sequential Algorithm
(described in Section~\ref{sec-wctt} above)
is non-uniform.

A preliminary look illustrates the nature of the problem.  For
example, consider the question of whether the USA is assigned to Group~A
with Qatar.  (This is an important question, since Qatar is weaker than
the other teams in Pot~1, so Group~A is the most desirable placement.)
Under the FIFA method, any of the 8 teams in Pot~2 is equally likely to be
selected first, and none of them have any regional conflict with Qatar, so
the USA has probability exactly 1/8 or 12.5\% of being placed in Group~A.
But under the uniform distribution, the USA should be {\it less} likely to
land in Group~A, because that leaves fewer ways for the numerous \Eu\
teams to be placed in the other groups.  (Indeed, we shall see below that
the uniform distribution gives a probability of approximately 9.0\% to the USA being
in Group~A, which is significantly smaller.) These calculations reinforce the
findings of \cite{csato2022unknown} Figure 1, which illustrated that the average strength of Group A is much smaller compared to the
 average strength of the other seven groups.

To compute the differences between the two distributions, we shall use a
Monte Carlo approach.  Thus, we will repeatedly simulate from both the
uniform distribution $\U$, and the distribution $\Pr$ of draws created by
the FIFA sequential method.  We now consider each of these problems in
turn, in Sections~\ref{sec-rej} and~\ref{sec-fifadraw} below.

Note that refinements on our basic Monte Carlo approach are possible, for instance we could utilise the exchangeability between teams in each pot from the {\em same} continental confederation. Thus for instance
we know that (for both $\U$ and $\Pr$) the probabilities of Mexico being placed in Group 1 with Qatar is identical to that for USA being placed in Group 1 with Qatar. Using these symmetries can improve the Monte Carlo accuracy. However in practice we found that we could achieve high levels of accuracy without needing these symmetries.

\subsect{Uniform simulation by rejection sampling}
\labelss{sec-rej}

To obtain draws with equal probabilities, the simplest way is to
use a Rejection Sampler.  This algorithm proceeds as follows:

\begin{enumerate}

\item First, assign a draw
completely randomly, without regard to the geographical constraints.
That is, select uniformly at random from the
$7! \times (8!)^3 \doteq 3.3 \times 10^{17}$ possible unconstrained draws
discussed in Section~\ref{sec-wctt} (which is straightforward).

\item Then, compute the number of teams from each region in each
group, and check if the geographic constraints are satisfied
(i.e., 1 or 2 \Eu\ teams plus 0 or 1 from each other region in each group).

\item If the constraints are satisfied, then output the chosen draw.

\item If the constraints are {\it not} satisfied, then reject the chosen
draw, and start again at step~1 above with a fresh unconstrained draw.

\end{enumerate}

\medskip
It is well-known (e.g.\ \cite[Section~II.3]{devroye}; see also
\cite{guyon2015preprint})
that this Rejection Sampler produces a draw
which is uniform over all possible valid draws, i.e.\ which is distributed
as the uniform distribution $\U$.  Then, by sampling many times and
averaging the results~\cite{drawprobs},
we can get a good estimate (by the Law of Large
Numbers) of the expected value according to $\U$
of any functional, or the probability
of any particular event.

In addition, our simulations~\cite{drawprobs} show that
about 1 in \ratio\ of the proposed draws were accepted,
i.e.\ satisfied the
geographical constraints.  This means that
the total number of valid draws is approximately
$7! \times (8!)^3 \bigm/ \ratio \doteq$ \NumbDraws.

\subsect{Simulating the FIFA sequential draw}
\labelss{sec-fifadraw}

Although the FIFA Sequential Draw method is easy to describe
(see Section~\ref{sec-wctt}), and can usually be implemented
without difficulty,
it is surprisingly challenging
to simulate with a computer program
(and neither FIFA nor UEFA provides an algorithm for this).
This is because potential
group assignments have to be skipped whenever they would conflict
with the geographical constraints.

Now, immediate conflicts are simple to detect.  Indeed, if adding a team
to a group would create a third \Eu\ team, or a second team from any
other region, then obviously that group must be skipped.
However, subsequent conflicts are much more complicated.  It may be
that assigning a certain team to a certain group would then make it
impossible to fill in the rest of the draw in a valid way, either
due to being forced to later add a third \Eu\ team or second team from
another region to a group, or failing to add at least one \Eu\ team to a
group.  These subsequent conflicts are usually relatively simple to resolve
(and indeed, they did not occur at all in the actual FIFA 2022 draw), but
this is not guaranteed.

For an extreme example, suppose a draw in
progress has teams from the various regions as follows:

\bigskip\noindent
\begin{tabular}{|l|c|c|c|c|c|c|c|c|c|}
\hline
{\bf \quad Group:} & A & B & C & D & E & F & G & H & To Go \cr
\hline
{\bf Pot~1} & \Af & \NA & \NA & \NA & \NA & \NA & \NA & \NA & \cr
\hline
{\bf Pot~2} & \As? &  &  &  &  &  &  &  & \SA,\SA,\SA,\SA,\SA,\SA,\SA\cr
\hline
{\bf Pot~3} & &  &  &  &  &  &  &  & \Eu,\Eu,\Eu,\Eu,\Eu,\Eu,\NA,\SA\cr
\hline
{\bf Pot~4} & &  &  &  &  &  &  &  & \Oc,\Oc,\Oc,\Oc,\Oc,\Oc,\Oc,\Oc\cr
\hline
\end{tabular}

\bigskip\noindent Then placing the \As\ team from Pot~2 into Group~A already
creates a subsequent conflict, since the seven \SA\ teams from Pot~2 must then
be placed in Groups~B through~H, and then in Pot~3 the \NA\ and
\SA\ teams cannot both be placed in valid groups without violating their
respective geographic constraints.
(Instead, the \As\ team in Pot~2 must be placed in some other group
besides Group~A, thus leaving e.g.\ Group~A for the \NA\ team from Pot~3,
and the other group for the \SA\ team from Pot~3.)
However, this subsequent conflict might
not become apparent until the end of the Pot~3 placements.

This example illustrates that a general program to simulate the FIFA
method must be robust enough to detect subsequent conflicts many steps
later.  Our approach~\cite{drawprobs} was to program this recursively.
We created a subroutine ``placerest()'' which, given a partial draw,
randomly selects a team to place next, and then
attempts to place that team in the next available slot.  It then
recursively calls itself with that one additional placement, to see if the
rest of the draw can then be filled in successfully.  If it can, then the
draw is complete.  If it cannot, then it instead attempts to place that
team in the subsequent available slot.  Continuing in this way, it
eventually successfully places every team.  The recursive nature of
the program ensures that any subsequent conflicts will be dealt with, no
matter how much further along they occur; for details see the computer
program at~\cite{drawprobs}.
(Related backtracking algorithms are presented in
\cite{guyon2015rethinking}, and in
\cite{csato2021fairness} which also makes connections to the
well-known computer science problem of generating all permutations
of a given sequence.)

Thus, this algorithm produces a draw with probabilities as in the current
FIFA sequential method, i.e.\ which is distributed as the FIFA
distribution $\Pr$.  Then, by sampling many times and averaging the
results~\cite{drawprobs}, we can get a good estimate
(by the Law of Large Numbers) of the expected value according to $\Pr$
of any functional, or the probability of any event.

\subsect{Specific probability comparisons}

Armed with our computer program~\cite{drawprobs}, we can now compare the
probabilities of various events according to the uniform distribution $\U$
and the FIFA Sequential Method $\Pr$.
We proceed by running each method one million times, taking the fraction of
successes as our point estimate, and also computing a corresponding
95\% confidence interval.

For example, the probability that England is in the same group as Germany
should be 10.53\% (95\%CI=10.47--10.59)
under $\U$, but increases to 11.78\% (95\%CI=11.72--11.85) under $\Pr$.  The
probability that Germany is in the same group as Qatar should be 13.74\%
(95\%CI=13.67--13.8)
under $\U$, but decreases to 12.50\% (exact) under $\Pr$.  The probability that
Canada is with Qatar should be 15.53\% (95\%CI=15.46--15.60)
under $\U$, but increases to 16.51\% (95\%CI=16.44--16.58)
under $\Pr$.  And, the probability that USA is with Qatar should be just
9.06\% (95\%CI=9.00--9.11)
under $\U$, but increases to 12.50\% (exact) under $\Pr$
(a relative increase of 38\%).
Probabilities of other events can similarly be computed~\cite{drawprobs}.

These probability differences are not huge, but they are large enough to
illustrate that the FIFA method is significantly different from a uniform
draw, and could lead to significantly different group assignments.
We now consider how to fix this problem, to achieve uniform group draws
while still preserving the interest and excitement of the FIFA method.

\sect{A motivating example}
\labels{sec-motiv}

To see more clearly why FIFA's sequential procedure
(described in Section~\ref{sec-wctt} above)
fails to achieve uniformity, and how we might correct that,
consider a simplified set up with only 6
teams to be allocated to 3 groups (of 2 teams each).
Suppose there are two seeded pots
(a subset of the actual World Cup 2022 pots):

{\bf Pot~1:} \ Qatar (Q) [\Af], France (F) [\Eu], Brazil (B) [\SA]

{\bf Pot~2:} \ Mexico (M) [\NA], Switzerland (S) [\Eu], Uruguay (U) [\SA]

\noindent
Qatar, as hosts, were automatically pre-assigned to Group~A.
Without loss of generality, assume that
France is placed in Group~B, and Brazil in Group~C.
Then the two \SA\ teams,
Brazil in Pot~1 and Uruguay in Pot~2, must be kept apart,
so Uruguay cannot be placed in Group~C.
There are thus four possible draws for assigning the three different groups:

$D_1$:\ QM, FU, BS.

$D_2$:\ QS, FU, BM.

$D_3$:\ QU, FM, BS.

$D_4$:\ QU, FS, BM.

\medskip
Let $\Pr$ be the probability measure resulting from FIFA's sequential
procedure described above (adapted to this simplified setting), and let
$\U$ be the probability measure which assigns equal probability to each
valid draw.
Thus $\U(D_i)=1/4$ for $i=1, 2, 3, 4$.
However, a simple calculation gives that
$\Pr(D_1)=\Pr(D_2)=1/3$ while $\Pr(D_3)=\Pr(D_4) = 1/6$.
In particular, if QU is the event that Qatar is paired with Uruguay, then
$\Pr(QU)=1/3$ whereas $\U(QU)=1/2$.
This clearly illustrates the potential non-uniformity arising from
FIFA's sequential procedure
(as also noted elsewhere, e.g.\ \cite[Footnote~19]{klossner2013odd}).

\subsect{In search of debiasing: random order sequential procedures}
\labelss{sec-multex}

It is natural to ask how to fix the biases of the above example while
retaining the sequential nature of the draw. 
Perhaps the most obvious try
is to randomise the order in which opponents for the Pot~1 teams are
found.
In certain cases, judicious choices of ordering and randomness can reduce bias (for example see \cite{csato2021fairer}) but here we shall investigate whether such a procedure can provide generic solutions to the problem.

In the usual FIFA method, the first drawn team in Pot~2
is first attempted to be placed in Group~A with Qatar.
To emphasise that we start with Qatar,
write $\Pr_Q$ for the probabilities induced by this method
(so $\Pr_Q=\Pr$ from above).
If instead the first drawn team in Pot~2 is attempted
to be placed in Group~B with France, then this leads to different
probabilities $\Pr_F$, and similarly $\Pr_B$ if the first team is
attempted to be placed in Group~C with Brazil.
Alternatively, if
the first drawn team in Pot~2 is attempted to be placed in Group~A
or~B or~C with probability $1/3$ each,
then such a procedure could be called a
{\em random order sequential draw},
with resulting
probabilities given by $\Pr_{rand} = {1 \over 3} (\Pr_Q+\Pr_F+\Pr_B)$.
(In principle, we could also specify that the
{\it second} team drawn from Pot~2 should have their attempted group
randomised among the two remaining teams, but
this example is sufficiently simple that we can ignore that.)

In this example, it can be checked that $\Pr_Q$ and $\Pr_F$ are both
non-uniform, but $\Pr_B$ happens to be uniform, i.e.\ $\Pr_B=\U$.
Moreover, a simple calculation yields that
$$
\Pr_{rand} \ := \ {1\over 3} (\Pr_Q+\Pr_F+\Pr_B )
\ = \ \U ,
$$
i.e.\ the random order sequential draw is uniform in this case.
This leads to the question of whether random order sequential draws are
always uniform.

To answer this question, consider a modification of our
simple example. Suppose that we
also forbid pairings between two \Eu\ teams, thus disallowing $D_4$
(which pairs France with Switzerland). In that case,
$$
\U(D_1) = \U(D_2) = \U(D_3) = 1/3.
$$
For this modified example,
$\P=\Pr_Q$ is easily seen to be uniform,
but $\Pr_F$ and $\Pr_B$ both turn out to be non-uniform.
Moreover, for the random order sequential draw
$\Pr_{rand} := {1\over 3} (\Pr_Q+\Pr_F+\Pr_B )$,
we calculate that
$$
 \Pr_{rand}(D_1)=5/18;\quad
 \Pr_{rand}(D_2)=13/36;\quad
 \Pr_{rand}(D_3)=13/36 \ .
$$
So, in this case, FIFA's fixed order draw $P_Q$ is
uniform, while the randomised order draw $P_{rand}$ is non-uniform.
This shows that randomised order draws do not solve the
non-uniformity problem. In fact, they can sometimes make the non-uniformity
worse, or even introduce non-uniformity in case where
the original FIFA algorithm happened to be uniform.
These findings in this top example concur with the findings of \cite{csato2021fairness} that
the order of pots can have a non-negligible effect on the size of the
bias in the draw mechanism, and that randomising the order cannot generically
solve the bias problem.

\subsect{In search of debiasing: multiple ball procedures}

We now return to the original example,
where the only constraint is on the two \SA\ teams.
Then we see from the above that two of the possible draws ($D_3$ and $D_4$)
put Uruguay in Group~A with Qatar, while only one ($D_1$) puts Mexico
in Group~A, and one ($D_2$) puts Switzerland in Group~A.  This suggests
that, to achieve the uniform distribution $\U$,
when selecting the team from Pot~2 to put in Group~A, we could use
a bowl with {\it two} Uruguay balls, and only one from each of Mexico and
Switzerland.
The next team in the draw is then selected uniformly at random
from those four balls, just as before.
This simple ``multiple balls solution'' thus achieves the correct
conditional probability (in terms of $\U$) for the team from Pot~2
to be placed in Group~A.

More formally, this solution could be described as follows.  When
selecting the team from Pot~2 to put in Group~A, we first count, for each
team in Pot~2, the number of valid draws which have that team in that
position (while keeping all of the
previously-selected teams in their previously-selected
positions, too).  In the above example, we have $n_U=2$ for Uruguay,
and $n_M=1$ for Mexico, and $n_S=1$ for Switzerland.
We then place $n_U$ balls for
Uruguay, $n_M$ balls for Mexico, and $n_S$ balls for Switzerland, all into a
bowl, and then sample one of the balls uniformly at random.
In this way, we select
Uruguay with probability $n_U/(n_U + n_D + n_S)$, and so on.
This ensures the correct conditional probability for the next position.
Hence, if this procedure is repeated for each new position, then it
ensures the correct full uniform probability $\U$ for the entire draw.

Can this procedure be extended to larger draws? In principle, yes.
However, the number of balls would soon get completely out of hand. For
instance in the World Cup 2022 draw, when choosing the Pot~2 team to play
Qatar, we would need to put about $(8!)^3 / 560 \doteq 10^{11}$
balls into the bowl, clearly impossible in practice.
Such pure multiple-ball draws
were already considered in~\cite{guyon2015rethinking},
which also noted their impracticality in larger draws.
Nevertheless, we shall see in
Section~\ref{sec-multballs} below that we can exhibit a practical and
completely uniform multiple balls solution to this problem, using
far smaller total numbers of balls.

\sect{A Metropolis (swap) algorithm solution}
\labels{sec-swap}

The Rejection Sampler algorithm of Section~\ref{sec-rej} provides a
perfectly uniform group draw distributed according to $\U$, so in some
sense it completely solves the problem.  However, use of this algorithm
would require the public to ``trust'' the computer to sample correctly,
and would not provide any drama or entertainment value during the draw.
So, we next consider ways to achieve a more interesting and entertaining
and transparent draw while still preserving the uniform probabilities $\U$.

One solution is as follows.   Begin with a Rejection Sampler uniform
sample as above. Then, repeatedly choose two teams at random from the same
pot, and ``swap'' their group assignments provided that swap does not
violate any of the geographical constraints.  (If the swap would violate
any constraint, then the group assignments are left unchanged.) The use of
a Metropolis algorithm like this was first suggested but not implemented
in the context of the UEFA Champions League knockout draw by
\cite{klossner2013odd}.

The validity of this Metropolis solution follows from:

\begin{proposition}
If we begin with a valid draw chosen from $\U$ (e.g.\ using a Rejection
Sampler as in Section~\ref{sec-rej}), and then repeatedly perform a
fixed number of swap moves as described above, then the distribution
of the draw remains equal to the uniform distribution $\U$.
\end{proposition}

\proof
First note that the proposed swaps are
symmetric, since making the same swap
twice is equivalent to not changing at all.
Furthermore, they are accepted if the swapped draw is still valid,
otherwise they are rejected.
It follows that
the swap moves correspond
to a Metropolis algorithm~\cite{Metropolis1953}
with stationary distribution $\U$.
Hence, $\U$ is a stationary distribution for the Markov
chain corresponding to this method, i.e.\ this method
induces a Markov chain Monte Carlo (MCMC) algorithm~\cite{handbook}
which preserves the distribution $\U$ as it runs.
Hence, the overall distribution of the group assignment remains
uniform no matter how many swap moves are attempted.
\qed

The swap moves used by this method can easily be performed manually, by
choosing pairs of teams by drawing balls from bowls, and checking directly
if any geographic constraints would be violated by swapping them.
Unlike the FIFA Sequential Draw of Section~\ref{sec-fifadraw}, there are
no subsequent conflicts, just immediate conflicts, so they can be easily checked.

As more swaps are performed, the overall group assignment continues to
change randomly, in unexpected and entertaining ways which could make for
an exciting spectacle.
(A somewhat related
exclusive Gibbs sampler for the UEFA Champions League draw
was proposed in \cite{klossner2013odd}.)
If desired, a large number of initial swaps could be performed quickly by
a computer. Then, a certain fixed number of final swaps could be performed
manually, by physically selecting balls from urns to determine which two
teams are selected next for possible swap. The final assignment would then
be whatever configuration remains after the final manual swap has been
performed.

An interactive simulation of this method applied to the 2022 FIFA World Cup is available at~\cite{drawweb}.

\sect{A multiple-balls solution}
\labels{sec-multballs}

We now investigate how to generalise the multiple-ball method introduced
in Section~\ref{sec-multex} to a full World Cup draw.

This method fills in the groups one team at a time. At each step, the
computer generates a collection of balls corresponding to all the teams
who could potentially occupy the next spot. One of those balls is then
chosen uniformly at random, and that team is placed in the next spot.
Once all spots are filled, it provides a complete group draw.

Recall that, for this method to generate the uniform distribution $\U$,
the next team needs to be selected from its correct conditional
probability according to $\U$.  That is, given a partial draw, if $n_j$ is
the number of ways of completing the draw which put team~$j$ in the next
position, then we should select team~$j$ with probability proportional to
$n_j$.

This approach immediately presents several challenges.  How could we
compute the $n_j$ values,
or at least the corresponding probabilities $p_j = n_j \bigm/ \sum_i n_i$?
And even if we knew the $p_j$, how could we sample
with probabilities proportional to them, in a practical way
without needing a massive collection of $\sum_i n_i$ different balls?
We first consider the second problem, of sampling with probabilities $p_j$
(Section~\ref{sec-ratsim}).
We then explain how it suffices to use {\it estimates} of the $p_j$,
so that they do not need to be computed analytically
(Section~\ref{sec-ratest}).
This leads to a
practical algorithm for conducting the draws
(Section~\ref{sec-ratalg}).

\subsect{Discrete random rational simulation}
\labelss{sec-ratsim}

\def\D{{\mathbf D}}

Consider the problem of simulating a discrete random event from $J$
possible values with given non-negative
probabilities $p_1, \ldots, p_J$ summing to~1. Call this
distribution $\D$.  We wish to sample from $\D$ using a {\em rational
simulation algorithm}, i.e.\ select some small non-negative integer number
$m_j$ of balls representing each value $j$, and then
draw each ball with probability $1/M$ where $M=\sum_{j=1}^J m_j$.

Now, if the $p_j$ were small integer multiples of each other, then this
would be easy.  For example, if $p_1=1/2$, $p_2=1/3$, and $p_3=1/6$, then
we could choose $m_1=3$, $m_2=2$, and $m_3=1$, and then drawing uniformly
at random from the $3+2+1=6$ balls would accomplish our task.  However,
we do not wish to
assume that the $p_j$ are rationally related, and we want the
total number of balls $M$ to remain moderate even if the ratios of
the $p_j$ have no simple fractional form.
We therefore propose the following algorithm.
(Our solution
uses the stratified sampling strategy common in Sequential Monte Carlo
\cite{kitagawa1996monte,fearnhead1998sequential}.
Other resampling strategies could also be used;
see \cite{douc2005comparison} for a discussion of options.)

\begin{enumerate}
\item
Let $M=\lceil \max\{ (1 / {p}_j) : {p}_j > 0 \} \rceil$ be
the ceiling of the reciprocals of the non-zero ${p}_j$.
\item Set $r_j = M \, {p}_j$, for $1 \le j \le J$.
(Note that our choice of $M$ ensures that $r_j \ge 1$
for each~$j$ with $p_j>0$,
which guarantees at least one ball of type $j$ below.)
\item
For each $1 \le j \le J$, place
$a_j := \lfloor r_j \rfloor $ balls of type $j$ into the bowl.
\item
Set $u_j = r_j - a_j$, and $v_j=\sum_{\ell=1}^j u_\ell$ for
$1\le j\le J$, with $v_0=0$. Also set $K=v_J$.
(Thus $K = \sum_j u_j = \sum_j r_j - \sum_j a_j = M - \sum_j a_j$
which is a non-negative integer.)
\item
Simulate independent uniform random variables
$U_1, \ldots ,U_K$ with $U_i \sim \Uniform[i-1,i)$.
\item
For each $1 \le j \le J$,
let $b_j = \# \{ i : U_i \in [v_{j-1}, v_j) \}$
be the number of random variables $U_i$
which lie in the interval $[v_{j-1}, v_j)$,
and add $b_j$ additional balls of type $j$ to the bowl.
(Note that we must have $0 \le b_j \le 2$,
and $\sum_j b_j = K = M - \sum_j a_j$.
Furthermore, the total number of balls of type $j$ is $m_j = a_j + b_j$,
so the total number of balls in the bowl is
$\sum_j m_j = \sum_j a_j + \sum_j b_j
= \sum_j a_j + (M - \sum_j a_j)
= M$.)
\item
Select a ball uniformly at random from the $M$ balls in the bowl.
\end{enumerate}

\begin{proposition}
\label{prop:multb}
Given a collection $p_1,\ldots,p_J$ of non-negative probabilities summing
to~1,
the above procedure selects a ball of type $j$ with probability $p_j$.
\end{proposition}

\proof
The interval $[v_{j-1},v_j)$ has length $v_j - v_{j-1} = u_j < 1$.
If it lies entirely inside an interval $[i-1,i)$, then
$\P(U_i \in [v_{j-1},v_j)) = u_j$,
so $\E(b_j) := \E\big[\#\{i : U_i \in [v_{j-1},v_j)\}\big] = u_j$.
Or, if there is an integer $i$ with $i-2 < v_{j-1} \le i-1 < v_j \le i$,
then $\P(U_{i-1} \in [v_{j-1},v_j)) = (i-1)-v_{j-1}$
and $\P(U_{i} \in [v_{j-1},v_j)) = v_j - (i-1)$,
so we still have $\E(b_j) := \E\big[\#\{i : U_i \in [v_{j-1},v_j)\}\big]
= (i-1)-v_{j-1} + v_j - (i-1) = u_j$.
Hence, in any case, $\E(b_j) = u_j$, whence
$\E(m_j) = a_j + u_j = a_j + (r_j-a_j) = r_j = M p_j$.
That is, the expected number of balls of type~$j$ is proportional to $p_j$.
Hence, the probability that a ball drawn uniformly at random will be
of type~$j$ is also proportional to $p_j$.
Then, since $\sum_j p_j = 1$, this probability is actually
equal to $p_j$, as claimed.
\qed

\subsect{Estimating the probabilities}
\labelss{sec-ratest}

To use the previous algorithm for group draws would require
that we know the conditional probabilities
$p_j = n_j \bigm/ \sum_i n_i$ for the next team to be chosen
in a partially-completed group draw, where $n_j$ is the number
of ways of completing the draw with team~$j$ in the next position.

Now, it might perhaps be possible to compute  $\{n_j\}$ directly.  We first
observe that the number of possible completions depends only on the
geographic region match-ups of the various teams, not on the actual team
names.  So, in the 2022 FIFA World Cup as described in
Section~\ref{sec-wctt}, Pot~1 can be distributed arbitrarily without
affecting the subsequent $n_j$.  Then, all we need to know about the Pot~2
teams is how many \Eu\ teams were put in the same group as a \Eu\ team
from Pot~1, whether the two \NA\ teams were put with \Eu\ or \As\ or \SA\
teams, and so on.  This leads to cascading combinatorial counts for the
numbers of ways of putting different regions into different positions in
the draw.  Then, multiplying by corresponding factorials gives the numbers
of ways of placing actual teams into the draw.

Such combinatorial problems can eventually be solved.  However, they
quickly become rather complicated and messy.  Furthermore, they have to be
re-computed for each possible partial draw, and the calculations are
entirely different for different regional distributions of the teams in
each pot.  So, this does not appear to be a feasible way of proceeding.

\def\phat{\widehat{p}}

Fortunately, there is a practical alternative.
Proposition~\ref{prop:multb} remains true if
the above algorithm instead used values $\phat_j$ which were unbiased
estimates of the $p_j$.  That is:

\begin{proposition}
\label{prop:estmultb}
Given a collection $\phat_1,\ldots,\phat_J$
of non-negative unbiased estimators summing to~1,
with $\E(\phat_j)=p_j$, if we replace $p_j$ by $\phat_j$ throughout in
the above procedure, then it will still
select a ball of type $j$ with probability $p_j$.
\end{proposition}

\proof
Conditional on the values of the estimators $\phat_j$,
the proof of Proposition~\ref{prop:multb} shows that
the probability of selecting a ball of type~$j$ is equal to $\phat_j$.
Hence, taking expectations over the $\phat_j$, it follows that the
probability that the algorithm using the estimators $\phat_j$ will
select a ball of type~$j$ is given by $\E(\phat_j) = p_j$, as claimed.
\qed

\begin{remark}
The approach of using an estimator $\phat_j$ for simulation, without
knowing the true value $p_j$ is the defining feature of a collection of
simulation techniques known as retrospective sampling, see
\cite{retrospect}.  However, our use of these methods here is rather
different from those in the literature which have focused largely on
Bayesian inference and exact diffusion simulation problems.
\end{remark}
\bigskip

The advantage of Proposition~\ref{prop:estmultb} is that
there is a natural way to estimate the $p_j$ in an unbiased
way: classical Monte Carlo samples.  That is, given a partial draw, we can
generate a large number $N$ of valid draw completions using a Rejection
Sampler similar to Section~\ref{sec-rej} above.  (In fact it is even easier,
since part of the draw is already chosen and does not need to be sampled.)
Then, we can estimate $p_j$ by the fraction of those valid draw
completions which have team~$j$ in the next position.  This gives a good
unbiased estimator $\phat_j$ of the true conditional probability $p_j$ that
team~$j$ would be placed in the next position according to the uniform
distribution $\U$.  That provides the final piece of the puzzle for us to
produce an effective Multiple-Balls uniform draw sample, which we now
present.

\subsect{Generating draws with multiple balls}
\labelss{sec-ratalg}

Combining the previous ideas together gives the following algorithm
for choosing the team in the next position of
a uniform group draw with distribution $\U$, by drawing
uniformly at random from a modest number of balls, as follows:

\begin{enumerate}
\item
Select a positive integer valued algorithm parameter $N$.
\item
Simulate $N$ different uniformly-distributed
completions of the current partial draw,
using a Rejection Sampler.
\item
\label{step-phat}
For each team~$j$, let $n_j$ be the number of such completions
which have team~$j$ in the next position,
and set $\phat_j = n_j / N$.

\item
Let $M=\lceil \max\{ (1 / {\phat}_j) : {\phat}_j > 0 \} \rceil$ be
the ceiling of the reciprocals of the non-zero ${\phat}_j$.
\item Set $r_j = M \, {\phat}_j$, for $1 \le j \le J$.
(So, $r_j \ge 1$ whenever $n_j>0$.)
\item
For each $1 \le j \le J$, place
$a_j := \lfloor r_j \rfloor $ balls of type $j$ into the bowl.
\item
Set $u_j = r_j - a_j$, and $v_j=\sum_{\ell=1}^j u_\ell$ for
$1\le j\le J$, with $v_0=0$. Also set $K=v_J$.
\item
Simulate independent uniform random variables
$U_1, \ldots ,U_K$ with $U_i \sim \Uniform[i-1,i)$.
\item
\label{step-m}
For each $1 \le j \le J$,
let $b_j = \# \{ i : U_i \in [v_{j-1}, v_j) \}$
be the number of random variables $U_i$
which lie in the interval $[v_{j-1}, v_j)$,
and add $b_j$ additional balls of type $j$ to the bowl
(so the total number of balls of type $j$ is $m_j = a_j + b_j$).
\item
Select a ball uniformly at random from the $M$ balls in the bowl.
\end{enumerate}
{\bf Return to motivating example of Section \ref{sec-motiv}.}
Let us give  a simple illustration based on the example in Section \ref{sec-motiv}. Although in this example we can easily compute the true $p_i$ probabilities, we shall assume that we cannot so still need to carry out a Monte Carlo experiment to estimate them.
Let us assume that we have allocated Qatar to Group 1, France to Group 2 and Brazil to Group 3. Now to be consistent with the notation above, we (arbitrarily) number the Pot 2 teams: (Mexico, Switzerland, Uruguay) $= (1, 2, 3)$. So allocating the Pot 2 team to Group A, we have
$$
p_1=1/4;\ \ \ \ 
p_2 = 1/4;\ \ \ \ 
p_3 = 1/2\ .
$$
Suppose we pick $N=100$, and then carry out 100 simulations from a multinomial with the above probabilities obtaining
$(24, 29, 47)$ observations of each type. So $({\hat p}_1, {\hat p}_2, {\hat p}_3) = (0.24, 0.27, 0.47)$ and we we find that the largest value of $1/{\hat p}_i$ is $1/0.24$ = 4.167. The algorithm therefore dictates that we choose $M$ to be $5$. So we shall have 5 balls, but we haven't chosen what type they should be yet.
$M({\hat p}_1, {\hat p}_2, {\hat p}_3)= ({\bf 1}.2, {\bf 1}.45, {\bf 2}.35)$ so that we shall initially allocate $(1, 1, 2)$ balls of type 1, 2, 3 respectively, and assign the final ball according to the remainder probabilities $(0.2, 0.45, 0.35)$.

\medskip
Returning to the method in general, the validity of this method is ensure through the following result. It follows from Proposition~\ref{prop:estmultb} that:

\begin{proposition}
\label{prop:multbalg}
If the above procedure is used sequentially for each position of a group
draw, then the final group draw will have distribution $\U$, i.e.\ will
be equally likely to be any of the potential valid draws.
\end{proposition}

An interactive simulation of this group draw generation
method for the 2022 FIFA World Cup is available at~\cite{drawweb}.

\medskip
\begin{remark}
The above procedure can be improved in a few minor ways.
For example, the true conditional probabilities $p_j$ must be
the same for all teams in the same geographical region, so it is possible
after step~\ref{step-phat} to replace each $\phat_j$ by the average
of the $\phat_i$ over all teams in the same region as team~$j$.
Also, if it happens after step~\ref{step-m}
that the final numbers of balls $m_j$
have a non-trivial common factor, i.e.\
$gcd\{m_j\} > 1$, then each $m_j$ can be divided by this common factor
to produce a simpler draw which still maintains the same probabilities.
\end{remark}

\subsect{Managing $M$}
\labelss{sec-multmit}

From a practical point of view, it is important to know how large $M$ could be. The sporting authority could be severely embarrassed if the ran out of balls for a particular team! Moreover the practical feasibility of the draw depends on $M$ not being too large. The procedure described above actually will require a random number $M$ of balls. We return to our motivating example to illustrate.

\noindent
{\bf Return to motivating example of Section \ref{sec-motiv}.}
We recall that we wanted to carry out a Monte Carlo simulation of $N=100$ draws from a trivariate random variable with probabilities
$$
p_1=1/4;\ \ \ \ 
p_2 = 1/4;\ \ \ \ 
p_3 = 1/2\ .
$$
Suppose we obtained (unusual) counts such as $(3, 36, 61)$.  This would lead to a choice of $M=34$. In fact it is easy to see that the maximum possible $M$ value is $100$ in this case and $1/N$ in general. Fortunately this is vanishingly unlikely, and of course by taking $N$ sufficiently large, we can ensure that $M$
 is small with high probability. For instance for $N=100$ we can directly compute that $\Pr (M\le 6)$ to be around $0.042$ while 
 for $N=1000$ that probability decays to less than $10^{-11}$.

Unfortunately, the required $N$ for making the probability that $M$ is large depends crucially on the probabilities $\{p_i\}$.
For instance, suppose $p_1$ was (the smallest probability at) $0.1$. Then $M$ will usually have to be at least $10$ and we require $N=1000$ to ensure that $\Pr (M\le 15)$ to be around $10^{-4}$.

In the case of  the FIFA World Cup draw, we have the additional complication that we do not actually know the $p_i$s so cannot predict the size of $N$ required to make $\Pr (M\le k)$ sufficiently small. Empirically we have observed in simulations that the maximum $M$ value is extremely rarely above 16, and for any future draw we could carry out a similar empirical study. However this does not provide any guarantees.

There is however a modification of the solution above which allows us to specify a maximum possible $M$, say $M_{max}$. Once we have specified  $M_{max}$, we replace step 4 in the above algorithm by
\begin{description}
\item{4'.}
Let $M=\max \{ M_{max}, \lceil \max\{ (1 / {\phat}_j) : {\phat}_j > 0 \} \rceil \}$.
\end{description}
We then still set $r_j = M \, {\phat}_j$ for each $j$ as before.
This guarantees that we will never use a total of more than $M_{max}$ balls,
and the method remains completely valid.
The only disadvantage of this method is that
now it is possible to get $r_j=0$ balls even though $n_j>0$, though
this will only occur if $1/\phat_j > M_{max}$ which is quite unlikely
for reasonable choices of $M_{max}$.

One special case of this algorithm is when $M_{max}=1$. In this case, only a single ball is produced, and all the random simulation takes place within the computer, which is clearly undesirable.
In practice, the user should choose $M_{max}$ judiciously: not too large (for reasons of practicality) and not too small (for transparency and to avoid $r_j=0$),
e.g.\ $M_{max}=20$.

\sect{A multiple-rejections solution}
\labels{sec-multirej}

Finally, we present a somewhat different solution, which still selects the
teams sequentially one position at a time, and still uses uniform draws from
bowls with modest-sized numbers of balls representing the different teams,
and still produces a completely uniform valid draw having distribution
$\U$, but with details which are somewhat different as we now describe.

Given a partial draw,
suppose we know the number $n_j$ of valid ways of completing
the draw with team~$j$ in the next position.  In terms of this,
let $\T$ be the set of all teams with $n_j>0$,
and set $n_{max}=\max_{j\in\T} n_j$. Then, for each team~$j\in\T$,
we sample a geometric random variable
$G_j \sim$ Geometric$(n_j/n_{max})$.
(The $G_j$ could be produced automatically by a computer in advance.)
Thus, $G_j \ge 1$.
Also, if $n_j = n_{max}$, then $G_j=1$.
Furthermore, if the $n_j$ are all roughly equal, then
most (if not all) of the $G_j$ will be equal to~$1$.


Given these $G_j$ values, the selection process proceeds as follows:

\begin{enumerate}
\item
\label{step-chooseteam}
Choose one of the teams~$j\in\T$, uniformly at random
(e.g.\ from balls in a bowl).
\item
If team~$j$ has now been chosen a total of $G_j$ times,
then select team~$j$ for the next position in the draw.
\item
Otherwise, return to step~\ref{step-chooseteam} and again
choose a team uniformly at random.
\end{enumerate}

\noindent In this way,
the above procedure produces a ``race'' in which the different teams are
each trying to be the first to be chosen $G_j$ times.
Eventually one team will be chosen $G_j$ times,
and will then be selected for the next position in the draw.

The usefulness of this procedure is given by:

\begin{proposition}
\label{prop:lives}
If the above procedure is used sequentially for each position of a group
draw, then the final group draw will have distribution $\U$, i.e.\ will
be equally likely to be any of the potential valid draws.
\end{proposition}

\proof
It suffices to show that each new position of the draw is filled with the
correct conditional probability according to $\U$.  This follows because
the above procedure is actually
a way of carrying out a corresponding Rejection Sampler.  Indeed,
consider a Rejection Sampler with proposal distribution which is
uniform on $\T$, with acceptance probability $n_j/n_{max}$ for each
team~$j$.  Then each choice of team in step~1 corresponds to one proposal
from the Rejection Sampler.  And, $G_j$ corresponds to the number of times
that team~$j$ must be proposed before it is finally accepted.  Hence,
team~$j$ being the first to be chosen $G_j$ times is precisely equivalent
to team~$j$ being the team selected by a Rejection Sampler with target
probabilities proportional to $n_j/n_{max}$, i.e.\
proportional to $n_j$.  The result follows.
\qed

Although this multiple-rejections solution produces a valid uniform draw,
it requires knowledge of the actual $n_j$ values, which might be
difficult in practice (as discussed in Section~\ref{sec-ratest} above).
Moreover, allocating a team to a particular slot takes multiple draws so is more time-consuming, and therefore less practical.
Hence, for actual group draws, we believe that
the methods of Sections~\ref{sec-swap}
and~\ref{sec-multballs} above are preferable.

\sect{Informal Descriptions}
\labels{sec-informal}

In this section, we provide less formal/technical descriptions of our
proposed draw methods, to explain to less technical readers how they would
work in actual implementation.  Although we have presented a number of
possible methods (including Rejection Sampler, Multiple-Rejections, etc.),
here we focus on our two main practical methods, namely
the Metropolis (swap) method of Section~\ref{sec-swap},
and the Multiple Balls method of Section~\ref{sec-multballs}.
Both methods are also demonstrated interactively at~\cite{drawweb}.

\subsect{The Metropolis (swap) method}

This method begins with some valid initial assignment,
i.e.\ by placing all teams in some group in a valid way which satisfies
all of the restrictions.  This initial assignment could be done by
computer, or mechanically by any valid method, since it will be
significantly modified in any case.

Then, the method does some pre-specified number of ``swap''
moves.  Each swap move proceeds as follows:

\begin{itemize}
\item First, one of the
four pots is chosen, e.g.\ Pot~2.  This choice could be done systematically,
i.e.\ first choose Pot~1, then Pot~2, etc.
\item Then, two teams are
selected from that pot, e.g.\ Germany and Mexico.
This choice could be done by placing one ball in a bowl for each of the eight
teams in that pot, and then randomly selecting two of the balls.
(As an exception, if Pot~1 is chosen, then the host team should {\it not}
have a ball since they should always remain in Group~A.)
\item Then, check if it is possible to swap the two
selected teams, e.g.\ if the draw would still be valid if Germany and
Mexico switch to each other's group.  If it is possible, then the swap
should be made.  If not, then no change should be made.
\end{itemize}

A reasonable number of these swap moves should be performed.
We recommend first performing a large number (e.g.~50) on a computer,
which can be done quickly and easily.  Then, the final number (e.g.~20)
could be performed by celebrities selecting balls from bowls as above,
under the watchful gaze of an excited general public.

Once the pre-specified total number of swap moves have been performed, then
the resulting draw shall be the final draw for the tournament.
(For an interactive demonstration, click the ``Swap speed''
buttons at~\cite{drawweb}.)

\subsect{The Multiple Balls method}

This method fills in the spots on the draw sequentially.
The Pot~1 team in Group~A is always the host team, as usual.
Then, a team from Pot~1 is chosen for Group~B, then one for Group~C, and
so on.
Once the Pot~1 teams are all assigned,
then a team from Pot~2 is chosen for Group~A, then one for Group~B, and
so on.
Then, a team from Pot~3 is chosen for Group~A, then Group~B, and so on.
Finally, a team from Pot~4 is chosen for Group~A, then Group~B, and so on.

Each of the individual choices is conducted as follows.
The computer will specify which teams are eligible for the next spot,
and how many balls each of those teams should get.
Then, that collection of balls is placed into a bowl, and a single
ball is selected.  Whatever team is specified by that ball is then
chosen and placed in the corresponding position on the draw.

For example, when selecting the Pot~2 team to go in Group~E,
the computer might specify two balls for each of
Netherland and Uruguay, and one each for Denmark and Mexico, for six in total.
Then, those six balls are placed in a bowl, and one selected.
If the selected ball says Uruguay, then Uruguay is placed into Group~E.

Once each spot has been chosen, i.e.\ a team from each of the four pots
has been assigned to each of the eight groups, then the draw is complete,
and is taken as the final draw.
And, since the computer only allowed balls for valid choices of teams,
this complete draw will automatically be a valid draw.
(For an interactive demonstration, click the ``Balls Update''
button at~\cite{drawweb}.)

\sect{Discussion}
\labels{sec-disc}

This paper has considered the challenge of designing football
group draw mechanisms
which have the uniform distribution over all valid draw assignments,
but are also entertaining, practical, and transparent.
We have explained
(Section~\ref{sec-compare}) how to simulate the FIFA Sequential Draw method,
to compute the non-uniformity of its draws by comparison to a
uniform Rejection Sampler.
We have then proposed two practical methods
(Metropolis swaps in Section~\ref{sec-swap}, and Multiple Balls
in Section~\ref{sec-multballs})
of achieving the uniform distribution while still using balls and bowls in
a way which is suitable for a televised draw.
These two solutions can be tried interactively for the World Cup groups
at~\cite{drawweb}.

An important advantages of our methodology is its generality, allowing it to be immediately translated for use
with different geographic constraints and other restrictions. The methodology can also be used seamlessly in conjunction with the constraints proposed in \cite{csato2022incentive} for mitigating the risks of {\em tanking}, ie teams throwing games.
Indeed, given any definition of what constitutes a ``valid'' draw,
our methods can be used to simulate uniformly from all such valid draws,
in ways which are still entertaining, practical, and transparent.

We have focused here on methods which are practical, unbiased, respect the sequential nature which FIFA and other footballing authorities seem to require, and try to ensure that  in some sense {\em most} of the randomness takes place in a transparent way within a live television show. As detailed earlier, other literature has concentrated more on  changing draw mechanisms to try to ensure a balanced draw. Clearly there is a tension between ensuring balanced and creating an unbiased fair draw, and which aspect is more important will depend on political considerations.  However 
we note that the flexibility of our methods are readily transportable so that, should the authorities wish to impose further constraints to ensure balance, our methods can be readily translated to that context.

Our approach also has implications for other football draws, such as the
upcoming UEFA (European) Champions League group stage draw in August 2022.  For this draw UEFA uses a slightly
different procedure from FIFA, choosing first a club and then randomising uniformly over possible valid allocations for that club
(as opposed to choosing the lexicographically first available slot). In fact the UEFA Champions League last 16 draw in December 2000 was different again. In that case, slots were filled in order with only the possible teams being put into a glass ball from which one was drawn uniformly at random. Both of these UEFA procedures are biased, although it is an open question as to which mechanism gives the least bias. However all these scenarios are amenable to correction using the methods we have presented in this paper.


The UEFA procedure Champions League last 16 draw mechanism can be seen as an approximation to a
Sequential Monte Carlo algorithm (e.g.~\cite{doucet2001sequential})
in which all of the non-zero incremental particle weights
are assumed to be equal.
Also, the 2026 World Cup will have 48 teams instead of 32, leading to
different constraints and challenges.
We plan to study the UEFA and World Cup 2026 draws
in greater detail in subsequent work.

Our three proposed solutions are all practical and unbiased. However they all have disadvantages. 
Existing draw mechanisms for both FIFA and UEFA use background computer calculations. Our proposed methods are no different. However unlike existing methods, we do carry out random simulations in these calculations. This could be seen as a disadvantage as this randomness is not {\em transparent}. Existing draw mechanisms also use computer-aided decision-making to check feasibility. Since these decision can be checked a posteriori, these methods can thus be considered to be more transparent in some sense. On the other hand, such methods are clearly biased, so it seems the choice is between correctness and some degree of transparency.

Our solutions all balance transparent randomness (i.e.\ celebrities drawing balls from glass bowls) with hidden randomness (i.e.\ computational calculations carried out instantaneously within the background to determine, for instance, the number of teach type of balls to use, or the number of lives a team possesses). The aim is to carry out as much of the draw from transparent randomness as possible, and all three methods can do this reasonably effectively. However it will be difficult to explain to a lay audience why some teams have larger number of balls (or lives) than others. Therefore both the multiple-ball and multiple-rejections solutions intrinsically require non-symmetry to the transparent randomness part of the draw. On the other hand the Metropolis (swap) solution, has the disadvantage of being very different from current procedures. Viewers might also be worried about apparent biases created by the draw obtained when the hidden computer random iterations have been completed, thinking {\em we started in the group of death, so it's no surprise that we ended up with a tough draw!}   

Note that in the Metropolis solution, we can in some sense decrease the proportion of computer randomness by running the final iterations carried out by balls and glass bowls for more iterations. Markov chain convergence arguments cound in principle also be used to quantify this iteration/transparency tradeoff.

Although practical, all our solutions are more complex than existing methods. For the multiple ball draw, when the number of balls required is large, the draw could become complex to implement. However see Subsection \ref{sec-multmit} for a discussion on how these risks can be mitigated or even completely removed. It should also be emphasised that no specialist computer or statistical expertise is required by those carrying out the draw. It is also not necessary for technical nomenclature such as geometric, binomial distributions or retrospective simulation techniques to be used at all as part of the draw.

In summary, we have proposed three solutions to solve the bias caused by the sequential draw mechanisms used by football authorities. These solutions are flexible and can be used with a wide variety of constraints and structures.
 We believe our methods could be used as they are or in modified form by football authorities to provided fairer draws for future tournaments, and we encourage these organisations to consider them seriously.

\bigskip
\bigskip

\ifanon

\newpartitle{Acknowledgements}
We thank the anonymous referees for detailed helpful reports.

\vskip 2cm

\else

\newpartitle{Acknowledgements}
We thank David Firth for helpful discussions of these topics,
and thank the anonymous referees for detailed helpful reports.
GOR was supported by EPSRC grants Bayes for Health (R018561) and CoSInES (R034710).
JSR was supported by NSERC discovery grant RGPIN-2019-04142.

\fi  

\bigskip
\raggedright
\bibliographystyle{plain}
\bibliography{fdraw}

\end{document}